\documentstyle[pra,eqsecnum,aps,amsmath,amssymb]{revtex}
\def\>{\rangle}
\def\<{\langle}
\def\n{\nonumber}
\def\lb{\left[}
\def\rb{\right]}
\def\lk{\left\{}
\def\rk{\right\}}
\def\<{\langle}
\def\>{\rangle}
\def\be#1\ee{\begin{equation}#1\end{equation}}
\def\ba{\begin{eqnarray}}
\def\ea{\end{eqnarray}}

\input{psfig}

\begin{document}
\draft
\twocolumn

\title{Distribution of local entropy in the Hilbert space of
bi-partite quantum systems: Origin of Jaynes' principle}
\author{Jochen Gemmer and G\"unter Mahler}
\address{Institut f\"ur Theoretische Physik,\\Universit\"at Stuttgart, Pfaffenwaldring 57,\\70550 Stuttgart, Germany}
\date{\today}
\maketitle


\begin{abstract}
For a closed bi-partite quantum system partitioned into system proper
and environment we interprete the microcanonical and the
canonical condition as constraints for the interaction between those
two subsystems. In both cases the possible pure-state
trajectories are confined to certain regions in Hilbert space. We show
that in a properly defined thermodynamical limit almost all states
within those accessible regions represent states of some maximum local
entropy. For the microcanonical condition this dominant state still
depends on the initial state; for the canonical condition it coincides
with that defined by Jaynes' principle. It is these states which
thermodynamical systems should generically evolve into.
\end{abstract}

\narrowtext

\section{Introduction}
Thermodynamics has been developed at a time, when the atomistic nature
of matter was not well understood. Statistical mechanics tried to fill
this gap, when, in turn, quantum mechanics as the micro-theory proper
was unknown yet. Meanwhile quantum theory has envoked many changes,
but the basic structure of modern theory is still close to that
formulated by Gibbs some hundred years ago and therefore still plagued
by old problems. Examples are the significance of ergodicity
\cite{NEU30,BLA59,BER55,PEN70} the nature of irreversibility
\cite{PEN70,TER91}, the origin of the second law
\cite{PEN70,TER91,GEM01,BER94} and the relation between physics and
information \cite{ZUR94,BAL99,LAN61}. In spite of entropy being one of the most fundamental concepts in
Gibbsian thermodynamics,
there have been endless discussions about its nature and its precise
definition \cite{BLA59,MLI93}. As entropy (-change) is
thermodynamically measurable (for any individual system), one would
like to see its reduction to mechanics still to be based on
observables. But entropy remains an alien concept
within this frame: rather than to a point (state) it refers to a certain accessible volume in
phase space, which depends on given macroscopic constraints such as energy, volume,
etc. Nevertheless, this definition,
eventually based on ergodicity, seems indispensible to
calculate thermodynamical state functions.\\
 On the other hand, one
wants entropy to account also for the
irreversibility, which is absent in the underlying microscopic
equations of motion. Thus the concept of entropy appears to be related to
the lack of knowledge that arises during
an evolution too complicated to be traced in detail. Some sort of
coarse-graining \cite{BLA59} is often envoked in this context, entropy would
then be expected to increase up to the limits set by those macroscopic
constraints. Such an information-theoretical approach also underlies
Jaynes' principle \cite{JAY57,JAYWS}. This principle aims to guide our
reasoning in the face of incomplete knowledge (``unbiased guess''). It
is not at all restricted to physics. But
from this point of view it seems as if entropy and the state functions
would depend on the physicist's ability to minimize his lack of
knowledege, which is, of course, unsatisfactory.\\
Within classical mechanics lack of knowledge may always be considered subjective: in
principle any observable can be known with unlimited precision. This is
different in quantum mechanics. From the uncertainty principle we know
that there are always observables that are undetermined. Nevertheless, in single system
scenarios, at least one observable can, in principle, be known exactly (pure states),
while for compound systems there are even states where all observables
refering to a specific subsystem are unknown, even
if some compound observable of the full system was exactly
predictable (pure but entangled and thus locally mixed states \cite{BAL99,BLU81}). Compound
systems might evolve from states that contain exact kowledge about
some observable of each subsystem (pure product states) into the above
mentioned states, featuring this fundamental lack of knowledge \cite{GEM02}.\\
So, in the quantum domain we have two possible sources of ignorance:
one being due to our inability to identify the initial state and calculate the
evolution exactly, the other being intrinsic to the momentary state
and thus present even for an``infinitely smart demon''.\\
Here we want to show that in typical thermodynamical situations the
fundamental lack of knowledge by far dominates over the subjective lack of
knowledge in
the following sense: Almost all the possible evolutions (of which we
are typically
unable to predict the actual one), will eventually lead to
states that are characterized by a maximum fundamental lack of
knowledge about the considered subsystem; this lack is only limited by
macroscopic constraints. Our considerations will entirely be based on 
quantum-mechanical system-environment-scenarios \cite{TER91,ZUR94,TAS98}.\\

\section{State measures and their averages}

\subsection{Definitions}

The entropy we are going to consider is the von Neumann entropy \cite
{BLU81} given by
\be
S(\hat{\rho}):=-k\mbox{Tr}\lk\hat{\rho}\ln \hat{\rho}\rk.
\ee
which is invariant with respect to any unitary transformation. In a diagonal representation of $\hat{\rho}$ this reads
\be
S(\{W_n\}):=-k\sum_n W_n \ln W_n,
\ee
where the $W_n$'s are the eigenvalues of $\hat{\rho}$, which are
routinely interpreted as the probabilities to find the system in the
eigenstate $|n\>$. The von Neumann
entropy is obviously related to the Shannon entropy. 
Another measure for the broadness of the probability distribution is
the purity $P$ \cite {TSA88}:
\be
P(\{W_n\}):=\sum_n  W_n^2 \quad \mbox{or} \quad P(\hat{\rho}):=\mbox{Tr}\lk\hat{\rho}^2\rk.
\ee
which is also invariant with respect to unitary transformations. In general, these two measures do not uniquely map onto each
other, but, as will be seen below, under some aditional conditions
maximum entropy states correspond to minimum purity states. Because of
its less complicated structure, we mainly consider $P$ and get back to
$S$ in the end.\\
Since we are going to deal with bi-partite systems partitioned into
the considered system, labeled by $g$ (``gas'') and the surrounding
labeled by $c$ (``container'') the pure state of the full system
will be denoted as
\be 
|\psi\>=\sum_{A,B}\sum_{a,b}\psi_{ab}^{AB}|A,a\>\otimes|B,b\>
\ee
where $|A,a\>$($|B,b\>$) denotes the $a$-th($b$-th) energy eigenstate with the energy
eigenvalue $E_A$($E_B$) of the gas (container) system, with
$a=1,2...N_A^g,b=1,2...N_B^c$. $N_A^g(N_B^c)$ are the respective
degrees of degeneracy.\\
From any such wavefunction the reduced density operator of the
subsystem, $g$,
is found by tracing over the container system:
\be
\hat{\rho}^g:=\sum_{A,A',B}\sum_{a,a',b}\psi_{ab}^{AB}(\psi_{a'b}^{A'B})^*|A,a\>\<A',a'|
\ee
For this state the purity is:
\be
\label{puri}
P^g=\sum_{ABCD}\sum_{abcd}\psi_{ab}^{AB}(\psi_{cb}^{CB})^*\psi_{cd}^{CD}(\psi_{ad}^{AD})^*
\ee
Here and in the following $a,c,A,C,$ label the gas, $b,d,B,D,$ the
container subsystem. (Note that $P^g=P^c$ as the total state is taken
to be pure) \\ 

\subsection{Time and path averages}

Now, let us consider the time average of some state measure $M$ which
could be the purity, the entropy, etc.
\be
\label{taver}
\overline{M}:=\frac{1}{T}\int_0^T M(|\psi(t)\>)dt
\ee
Choosing a special parametrization for $|\psi\>$, we can convert the time
integral into an integral over the trajectory generated by the total
system's Hamiltonian for given initial conditions. Parametrizing
$|\psi\>$ in terms of the real ($\psi_i$) and imaginary ($\psi_i'$) parts of
its amplitudes in some basis $\{|i\>\}$ we can write instead of (\ref{taver})
\be
\label{killer}
\overline{M}=\frac{\int_{|\psi(0)\>}^{|\psi(T)\>}M\left(\lk
  \psi_i,\psi'_i\rk\right)\frac{1}{v_{eff}} |d|\psi\>|}{\int_{|\psi(0)\>}^{|\psi(T)\>}\frac{1}{v_{eff}}|d|\psi\>|}
\ee
where $|d|\psi\>|$ denotes the ``length'' of an infinitesimal step along the
trajectory in Hilbert-space.\\
The advantage of this special
parametrization derives from the fact that the effective velocity
\be
\label{vef}
v^2_{eff}=\sum_{i}\left(\dot{\psi}_i^2+\dot{\psi '}_i^2\right)=\frac{1}{\hbar^2}\<\psi (0)|\hat{H}^2|\psi (0)\>
\ee
is constant on each trajectory and thus independent of the actual
time $t$ or the special point on the trajectory. Hence, the integral
  (\ref{killer}) simplifies to
\be
\label{patha}
\overline{M}=\frac{1}{L}\int_{|\psi(0)\>}^{|\psi(T)\>}M\left(\lk
  \psi_i,\psi'_i\rk\right)|d|\psi\>|
\ee
where $L$ is the length of the path. So, in this parametrization of Hilbert space, the time average of any state
  measure equals its path average along any given trajectory. We will
  exploit this relationship to infer the time-average (or even the
  ``typical'' momentay value) of $M$ for cases, in which we know that $M$ is
  constant over a large part of the Hilbert space region accessible by
  the respective trajectory, (see Sect.{\bf \ref{lan}}).   

\section{Microcanonical constraints}

\subsection{Accessible Hilbert space region}

Any thermodynamical system has to allow for a partition of its full
Hamiltonian in the following way:
\be 
\label{ther}
\hat{H}=:\hat{H}^g+\hat{H}^c+\hat{I}\quad \mbox{with} \quad
\<\hat{I}\>\ll\<\hat{H}^g\>,\<\hat{H}^c\>
\ee
where $\hat{H}^g$ and $\hat{H}^c$ are the local Hamiltonians of
the system and the environment, respectively, with
$\lb\hat{H}^g,\hat{H}^c\rb=0$, $\hat{I}$ is some sort of
interaction and the inequality for the expectation values has to hold for all
states that the system can evolve into under given constraints. If a partition according to this weak coupling scheme was impossible, the idea of
system proper and surrounding would be meaningless. Furthermore, weak
coupling is required to justify the concepts of extensive/intensive
variables, a basic feature of thermodynamics \cite{DIU89} (cf. also \cite{ALL00})\\
Making such a partition possible, though, might involve a re-organisation of the
Hamiltonian. Consider, for example, an ideal gas in a
container. $\hat{H}^g$ in this case is a free particle Hamiltonian,
$\hat{I}$ contains the interactions of all gas particles with all
container particles. In this case, the expectation value of $\hat{I}$
for an energy eigenstate of the uncoupled problem
$\hat{H}^g+\hat{H}^c$ would definitely not be small. But this deficiency can be overcome by defining an effective local Hamiltonian and an effective
interaction:
\be
\hat{H}^{g'}:=\hat{H}^g+\hat{V^g} \quad \hat{I}':=\hat{I}-\hat{V^g}
\ee
Here, $\hat{V^g}$ is some potential for the gas particles describing the
mean effect of all container particles together. Substituting the real
by the effective parts of the Hamiltonian obviously leaves the full
Hamiltonian unchanged, but now the partition fits into the above
scheme. In general, however, $\hat{I}'$ cannot be made zero.\\
It has often been claimed that a system under so-called microcanonical
conditions would not interact with its environment. This, however, is typically
not true (cf. \cite{BER55,BLA59}). A thermally isolated gas in a container,
e.g., definitely interacts with  the container, otherwise the gas could
not even have a well defined volume; the only constraint is, that this
interaction should not exchange energy. If the energies contained in the gas and the environment 
\be
E^{g'}:=\<\hat{H}^{g'}\> \quad E^c:=\<\hat{H}^c\>
\ee
are to be conserved, it follows that \cite{GEM01}
\be
\label{kommu}
\lb\hat{H}^{g'},\hat{I}'\rb=0 \quad \lb\hat{H}^c,\hat{I}'\rb=0 
\ee
Except for these constraints we need not specify $\hat{I}$ in more
detail. All interactions that fulfill this relation will create
perfectly microcanonical situations. Based on these commutator
relations we find that for any energy eigenspace specified by $A,B$
\be
\label{constr}
\sum_{a,b}|\psi_{ab}^{AB}(t)|^2=\sum_{a,b}|\psi_{ab}^{AB}(0)|^2
\ee
is a conserved quantity, set by the initial state. Since we want to consider cases here that have zero local entropy in the
beginning (product states), we get
\be
\label{nix}
\sum_{a,b}|\psi_{ab}^{AB}(0)|^2=\sum_{a,b}|\psi_a^A(0)|^2|\psi_b^B(0)|^2=W_A^gW_B^c
\ee
where $W_A^g\left(W_B^c\right)$ are the
probabilities of finding the gas-(container-) sytem somewhere in the possibly highly
degenerate subspace characterized by the energy
eigenvalues $E_A^g\left(E_B^c\right)$. This is the only constraint that
microcanonical conditions impose on the accessible region of Hilbert space.\\

\subsection{The ``landscape'' of $P^g$ in the accessible region}
\label{lan}

We are not able to compute the time average of $P^g$ according to (\ref{patha}) for we do not know $\hat{I}$ in detail, most likely we do
  not even know $\hat{L}^g$ and $\hat{L}^c$ in full detail, and even
  if we did, we could never hope to solve the Schr\"odinger equation for a
  system involving a macroscopic number of degrees of freedom. This is our subjective
  lack of knowledge. But, as will become clear shortly, this lack does
  not really matter.\\
To show this, we proceed as follows:\\
The minimum purity state consistent with the microcanonical conditions
  (\ref{constr}, \ref{nix})
  and its corresponding purity are:
\be
\label{pmin}
\hat{\rho}^g_{min}=\sum_{A,a} \frac{W_A^g}{N_A^g}|A,a\>\<A,a| \quad P^g_{min}=\sum_A\frac{(W_A^g)^2}{N_A^g}
\ee
We compute the average of $P^g$
over the total accessible Hilbert space region. We will show that this
average is very close to $P_{min}^g$ for a large set of
cases. Considering only these cases, we
can conclude that $P^g\approx P_{min}^g$ for almost all states within
this region. Since the only local state that has $P^g= P_{min}^g$ and is consistent
  with the microcanonical conditions is $\hat{\rho}^g_{min}$, all states
  within the accessible region that feature $P^g\approx P_{min}^g$ must yield reduced local states that are very close to
$\hat{\rho}^g_{min}$.\\
Furthermore, as there is no other constraint for the paths entering
(\ref{patha}) except to
  stay within the allowed region, any path will typically leave the
  very tiny region of maximum purity where it started if the initial
  state was a product state, and then venture through a region of
  almost minimum purity along almost all of its length. Thus in the long time limit
  any momentary state will be characterized by a maximum fundamental lack of knowledge.\\
To calculate the Hilbert space average of $P^g$ denoted as $<P^g>$ we
need a parametrization for $|\psi\>=\{\psi_{ab}^{AB}, \psi_{ab}^{AB'}\} $ confined to the allowed
region (\ref{constr}, \ref{nix}). The Hilbert space average can then be written as
\be
\label{hsa}
<P^g>=\frac{\int P^g\left(\lk\phi_n\rk\right)\det{\mathcal F}\prod_n d\phi_n}{\int\det{\mathcal F}\prod_n d\phi_n}
\ee
where $\phi_n$ is the respective set of parameters and ${\mathcal F}$ is the corresponding
functional matrix.\\
According to (\ref{constr}) and (\ref{nix}) the real and imaginary parts of the amplitudes that
correspond to a degeneracy subspace $AB$ must be parametrized to
lie on a hypersphere of radius $R=\sqrt{W^g_A W^c_B}$. Thus there is a
corresponding set of parameters $\lk\phi_n^{AB}\rk$ for each
degeneracy subspace on which the amplitudes of this and only this
subspace depend. This means that the functional matrix ${\mathcal F}$
has block form and its determinant factorizes such that
\be
\label{hsa}
<P^g>=\frac{\int P^g\prod _{AB}\det{\mathcal F}^{AB}\prod_n d\phi_n^{AB}}{\prod _{AB}\int \det{\mathcal F}^{AB}\prod_n d\phi_n^{AB}}
\ee
As a consequence, the average over each term of $P^g$ according to
(\ref{puri}) factorizes and
reduces to a product of averages over the degeneracy subspaces that
the amplitudes in that very term correspond to.\\

\subsection{ Hilbert space average of $P^g$ }
\label{int}

It turns out that the averages of the terms of $P^g$ belonging to the
same ``class'' have the same
functional dependence on the degree of degeneracy $N_{AB}:=N^g_AN^c_B$ of the
subspaces they correspond to and on
the probability of those subspaces to be occupied $W_{AB}:=W^g_AW^c_B.$ Thus, we first have to evaluate the
average of each term  within a given class, add all terms and finally sum over all classes.\\ 
There are six classes to be considered:\\
I: All four amplitudes correspond to different states $((A\neq C)\vee (a\neq c))\wedge((B\neq
D)\vee (b\neq d))$. Then each average is a product of averages of
products of different cartesian coordinates over full hyperspheres. As can be seen
from the appendix (\ref{sym}) those averages vanish.\\
II: The ``gas indices'' (indices refering to the gas subsystem) of all amplitudes correpond to the same
gas state, but the container indices to two different
container subspaces $((A=C)\wedge (a=c))\wedge (B\neq D)$. They factorize into: 
\ba
&&T^{II}=\left(\frac{\int |\psi_{ab}^{AB}|^2\det{\mathcal
F}^{AB}\prod_n d\phi_n^{AB}}{\int \det{\mathcal F}^{AB}\prod_n
d\phi_n^{AB}}\right)\times \n\\
&&\left(\frac{\int |\psi_{ad}^{AD}|^2\det{\mathcal F}^{AD}\prod_n d\phi_n^{AD}}{\int \det{\mathcal F}^{AD}\prod_n d\phi_n^{AD}}\right)
\ea
The application of (\ref{sym}) yields
\be
T^{II}=\frac{W_{AB}}{N_{AB}}\frac{W_{AD}}{N_{AD}}
\ee
There are $N^g_AN_BN^c_D$ terms belonging to that class and
subspace-combination. Since subspaces that have $B=D$ are excluded we find, summing
over subspaces:
\ba
&&\sum_{ABD}N^g_A(N^c_BN^c_D-\delta_{BD}(N_B^c)^2)T^{II}\\
&&=\sum_A\frac{(W^g_A)^2}{N^g_A}\left(1-\sum_B(W^c_B)^2\right)\n
\ea
III: The container indices of all amplitudes correspond to
the same container state, but the gas indices to two
different gas subspaces $(A\neq C)\wedge((B=D)\wedge (b=d))$. By
repeating the above procedure (II) we get:
\be
\sum T^{III}=\sum_A\frac{(W^c_B)^2}{N_B}\left(1-\sum_A(W^g_A)^2\right)
\ee
IV: The gas indices of all amplitudes correspond to the same
gas state and the container indices to the same container subspace but
to different container states within this subspace $((A=C)\wedge
(a=c))\wedge ((B=D)\wedge (b\neq d))$. Those terms are of the form
\be
T^{IV} =\frac{\int |\psi_{ab}^{AB}|^2|\psi_{ad}^{AB}|^2\det{\mathcal F}^{AB}\prod_n d\phi_n^{AB}}{\int \det{\mathcal F}^{AB}\prod_n d\phi_n^{AB}}
\ee
To evaluate this equation one needs (\ref{ku}) and gets:
\be
T^{IV}=\frac{(W^g_A)^2(W^c_B)^2 \Gamma (N_{AB})}{\Gamma
(N_{AB}+2)}=\frac{(W^g_A)^2(W^c_B)^2}{N_{AB}(N_{AB}+1)}
\ee
For each subspace-combination there are $N^g_AN^c_B(N^c_B-1)$ terms in that class. Thus, summing over subspaces we find:
\be
\sum_{AB}N_{AB}N^c_B(N^c_B-1)T^{IV}=\sum_{AB}\frac{(W^g_A)^2(W^c_B)^2(N^c_B-1)}{N_{AB}+1}
\ee
V: The container indices of all amplitudes correpond to the
same container state and the gas indices to the same gas subspace but
to different gas states within this subspace $((A=C)\wedge
(a\neq c))\wedge ((B=D)\wedge (b=d))$. For those terms the
above calculation (IV) has to be repeated yielding:
\be
\sum T^{V}=\sum_{AB}\frac{(W^g_A)^2(W^c_B)^2(N^g_A-1)}{N_{AB}+1}
\ee
VI: All four amplitudes correspond to the same
state. These terms read:
\be
T^{VI}=\frac{\int |\psi_{ab}^{AB}|^4|\det{\mathcal F}^{AB}\prod_n d\phi_n^{AB}}{\int \det{\mathcal F}^{AB}\prod_n d\phi_n^{AB}}
\ee
Using (\ref{ku}) one gets:
\be
T^{VI}=\frac{2(W^g_A)^2(W^c_B)^2 \Gamma (N_{AB})}{\Gamma
(N_{AB}+2)}=\frac{2(W^g_A)^2(W^c_B)^2}{N_{AB}(N_{AB}+1)}
\ee
Since there are $N_AN_B$ terms in this class for each subspace, one gets:
\be
\sum_{AB}T^{IV}N_{AB}=\sum_{AB}\frac{2(W^g_A)^2(W^c_B)^2}{N_{AB}+1}
\ee
If we now finally sum up all the contributions from the six classes we get:
\ba
\label{exact}
&&<P^g>=\\
&&\sum_A\frac{(W_A^g)^2}{N_A^g}\left(1-\sum_B(W_B^c)^2\right)+\sum_B\frac{(W_B^c)^2}{N_B^c}\left(1-\sum_A(W_A^g)^2\right)\n\\
&&+\sum_{A,B}\frac{(W_A^g)^2(W_B^c)^2(N_A^g+N_B^c)}{N_A^gN_B^c+1}\n
\ea
The Hilbert space average is thus a unique function of the invariants
$W_A^g,W_B^c$, specified by the initial product state, and the
degeneracies $N^g_A,N^c_B$.\\
If we, just to get in contact with previous results, ask for the average
purity of an arbitrary state with no constraints at all we can enlarge
the accessible region over all Hilbert space by formally taking
both subsystems as completely degenerate. Doing so, we recover
\be
<P^g>=\frac{N^g+N^c}{N^gN^c+1}
\ee
as a special case \cite{LUB78}.\\ 
If the degeneracy of the occupied energy levels is large
enough that $N_A^gN_B^c+1\approx N_A^gN_B^c$, which should hold true for typical thermodynamical systems,
(\ref{exact}) reduces to
\be
\label{apr}
<P^g>\approx\sum_A\frac{(W_A^g)^2}{N_A^g}+\sum_B\frac{(W_B^c)^2}{N_B^c}
\ee
The first sum in this expression is obviously exactly $P^g_{min}$
(\ref{pmin}), so that for systems and initial conditions, in which the
second sum is very small, the allowed region almost entirely consists of states for
wich $P^g\approx P^g_{min}$. The second sum will be small if the container
system occupies highly degenerate states, typical for thermodynamical
systems, in which the surrounding is much larger than the considered
system. This is the set of cases mentioned already in Sect. {\bf \ref{lan}}. Since the density operator that has $P^g=P^g_{min}$ and $S^g=S^g_{max}$
and is consistent with the microcanonical conditions is unique, the
density operators with $P^g\approx P^g_{min}$ should not deviate much
from this one and should therefore also have $S^g\approx S^g_{max}$. A
more detailed lengthy but
straightforward perturbative analysis shows that:
\be
\label{ende}
<S^g>\approx S^g_{max}\left(\lk W_A^g,N_A^g\rk
\right)-K\left(\sum_B\frac{(W_B^c)^2}{N_B^c}\right)
\ee
where $K$ is a
 positive function that scales linearly with the system size of the gas
 system. (\ref{ende}) is valid for the thermodynamical regime, i. e.,
if the second sum in (\ref{apr}) is small compared to the first one.

\section{Canonical constraints}

Our approach to the canonical conditions will be based on similar
techniques as before. The possibility of a partition according to
(\ref{ther}) is still assumed. But now there is no further constraint on the the interaction
$\hat{I}$, since energy is allowed to flow from one subsystem to the other. The only constraint for the accessible region therefore
derives from the initial state of the full system and the fact that
the probability to find the total system at some energy $E$
\be
\label{bed}
W_E:=\sum_{A,B/E}W_{AB}=\sum_{A,B/E}\sum_{a,b}|\psi_{ab}^{AB}|^2 
\ee
should be conserved. (Here $A,B/E$ stands for: all $A,B$ such that
$E_A+E_B=E$) One could try to repeat the above calculation under this
constraint, but now it turns out that the actual minimum purity is no longer
near the average purity
over the accessible region. Thus, one has to proceed in a
slightly different way.\\
Contrary to the microcanonical case the probability to find the
gas (container) subsystem at some given energy is no longer a constant of motion here. 
But we are going to prove that there is a predominant distribution, $\{W^d_{AB}\}$, applicable to almost all states within the allowed region. The
subregion formed by these states will be called the ``dominant
region''. Once the
existence of such a dominant region has been established, we can use the
results from the microcanonical conditions to argue that almost all
states within this dominant region feature the maximum local
entropy that is consistent with the predominant distribution.\\
Therefore, just like in the previous case, our subjective lack of
knowledge about where to find the system within the accessible region
should be irrelevant. The reduced local state $\hat{\rho}^g(t)$ as a function of the full state
$|\psi(t)\>$ should always evolve into a state with a fixed probability
distribution $W^g_A$ and an
almost time invariant entropy, which is the maximum entropy that is
consistent with this (canonical) distribution. Nevertheless, the state of the full
system continues to move with the constant velocity (\ref{vef}) in Hilbert space.

\subsection{Identification of the dominant region}

First, we calculate the size of a region in Hilbert space that is associated
with a certain energy distribution $\{W_{AB}\}$. This size will then be
maximized with respect to the $W_{AB}$'s under the condition of the
energy probability distribution of the whole system $\{W_E\}$
being kept fixed, in order to find the predominant distribution
$\{W^d_{AB}\}$. According to (\ref{hsa}) the size of the region asociated with the
energy distribution $\{W_{AB}\}$ is:
\be
A(\{W_{AB}\}):=\prod _{AB}\int \det{\mathcal F}^{AB}\prod_n d\phi_n^{AB}
\ee
Those integrals are just the surfaces of hyperspheres and can be done
using the techniques described in the appendix:
\be
A(\{W_{AB}\}):=\prod _{AB}W_{AB}^{N_{AB}-1/2}O(N_{AB})
\ee
Here $O(N_{AB})$ is the surface area of a $2N_{AB}$-dimensional
hypersphere of radius $R=1$.\\
Instead of maximizing $A$ directly, we choose to maximize $\ln A$; this
is equivalent, since the logarithm is a monotonous
function. Furthermore we drop all terms that do not depend on
$\{W_{AB}\}$ since they are of no relevance for the maximum and set
$N_{AB}-1/2\approx N_{AB}$, an approximation that is not
necessary but simplifies the calculation and is definitely valid for
large degrees of degeneracy. Introducing the Lagrange multipliers
$\{\lambda_E\}$, the function we want to maximize with respect to the
$W_{AB}$ reads:
\be
\ln \tilde{A}=\sum_{A,B}N_{AB}\ln W_{AB}-\sum_E \lambda_E
\left(\sum_{ A,B/E}W_{AB}-W_E\right)
\ee
This maximization is routinely done by solving the following set of
equations:
\be
\frac{\partial \ln \tilde{A}}{\partial W_{AB}}=0
\ee
and yields
\be
\label{domd1}
W^d_{AB}=\frac{N_{AB} }{\lambda_{E=E_A+E_B}}
\ee
Finally, using (\ref{bed}) we find for the Lagrange multipliers
\be
\label{domd2}
\lambda_E=\frac{N_E}{W_E} \quad N_E=\sum_{A,B/E}N_{AB}
\ee
We have thus identified the energy probability distribution, which most of the states within the
accessible region exhibit, i. e., the energy probability distribution of the
dominant region, $\{W^d_{AB}\}$.

\subsection{Analysis of the size of the dominant region}
 
So far we have only shown that among the regions with given energy
probability distribution $\{W_{AB}\}$  there is a biggest (or smallest) one. But for our argument we
need to show that this region $\tilde{A}^d$ is, indeed, extremely larger than
all the others, that it really fills almost the entire accessible
region. To examine the size of this region we need to know, how the
size of a region depends on the corresponding distribution
$\{W_{AB}\}$, if this distribution does not deviate much from the
dominant distribution $\{W^d_{AB}\}$. Therefore we consider
$W_{AB}=:W^d_{AB}+\epsilon_{AB}$, where the $\epsilon_{AB}$'s are
supposed to be small. For $\ln \tilde{A}$ we then find
\be
\ln \tilde{A}=\sum_E \sum_{A,B/E}N_{AB}\ln (W^d_{AB}+\epsilon_{AB})
\ee
with
\be
\label{null}
\sum_{A,B/E}\epsilon_{AB}=0
\ee
The latter condition guarantees that $W_{AB}=W^d_{AB}+\epsilon_{AB}$
still belongs to the accessible region.\\
Expanding the logarithm to second order we get:
\be
\ln \tilde{A}\approx \sum_E \sum_{A,B/E}N_{AB}\left(\ln
(W^d_{AB})+\frac{\epsilon_{AB}}{W^d_{AB}}-\frac{1}{2}\left(\frac{\epsilon_{AB}}{W^d_{AB}}\right)^2\right) 
\ee
Since the expansion is around an extremum the linear term should
vanish. Indeed, using (\ref{null}) the second summation over this term yields:
\be
\sum_{A,B/E}N_{AB}\frac{\epsilon_{AB}}{W^d_{AB}}=\sum_{A,B/E}\lambda_E\epsilon_{AB}=0
\ee
so that, using (\ref{domd1}) and (\ref{domd2}), we finally find:
\be
\label{doms}
\tilde{A}\approx
\tilde{A}^d\prod_{AB}\exp{-\frac{\left(\sum_{A,B/E}N_{AB}\right)^2}{2N_{AB}W_E}\epsilon_{AB}^2}
\ee
i.e. regions, $\tilde{A}$,  that
correspond to energy probability distributions that deviate from the dominant
one are smaller than the dominant region, $\tilde{A}^d$. Since the smallest
factor that can appear in the exponent of (\ref{doms}) for given $N_E$
is $\frac{N_E}{2W_E}$, the regions $\tilde{A}$ will be extremely
smaller already for very small deviations if the corresponding $N_E$'s
are large.

\subsection{The canonical distribution}

Finally, to find the marginal, dominant energy probability distribution $W^d_A$
of the gas system one has to sum the
compound probabilities $W^d_{AB}$ over the irrelevant container system
to obtain:
\be
\label{ha}
W^d_A=\sum_B W^d_{AB}=\sum_B \frac{N_{AB}}{\lambda_E}=N^g_A\sum_B \frac{N^c_BW_E}{N_E}
\ee
Going now from the degrees of degeneracy, $N^g_A, N^c_B$, to the state densities
$n^g(E^g),n^c(E^c)$ and from sums to integrals, (\ref{ha}) reads:
\be
\label{fast}
W^d(E^g)=n^g(E^g)\int_{E_A}^\infty\frac{W(E)}{n(E)}n^c(E-E^g)dE
\ee
To better appreciate the meaning of this result we first consider the case of
the full system energy being exactly specified in the beginning ($W(E)=\delta
(E-U)$). Then (\ref{fast}) reduces to:
\be
W^d(E^g)=\frac{n^c(U-E^g)n^g(E^g)}{n(U)}
\ee
This is precisely the form postulated in older approaches \cite{SCH89} by
simply taking the set of local energy eigenstates of both subsystems
for the full space of the system. In this way the authors excluded superpositions or
entangled states, but had to assume perfect ergodicity, an assumption that
is not needed in our approach. (\ref{fast}) is just a combination of
such distributions weighted by the probability to find the total
system at the corrsponding energy and then adequately normalized.\\
For the remainder we follow those older approaches and their assumptions
about the spectra of large systems. Thus we apply the standard expansion of $n^c(E-E_A)$ that is supposed
to be valid even for large $E^g$, if the system is large,
\be
n^c(E-E^g)\approx e^{\ln n^c(E)-\frac{d}{dE}\ln n^c(E)E^g}
\ee
so that we can write
\be
W^d(E_A)\approx
n^g(E^g)\int_{E_A}^\infty\frac{W(E)n^c(E)}{n(E)}e^{-\frac{d}{dE}\ln
n^c(E)E^g}dE
\ee
Assuming the function $\frac{W(E)n^c(E)}{n(E)}$ to be still peaked sharply enough at
some value $E=U$, such that $\frac{d}{dE}\ln n^c(E)$ does not vary much
in the range where the former function takes on non-negligible values, (and
again, this range can be fairly wide, if the system is large) we get 
\be
W^d(E_A)\propto n^g(E^g)e^{-\frac{E^g}{kT}}
\ee
where we have defined $\frac{d}{dU}\ln n^c(U):=\frac{1}{kT}$.\\
A set-up  similar to the one that we discussed to derive the canonical
distribution, but more spezialzed, has been analized by Tasaki
\cite{TAS98}. His results for which entirely
different techniques were used and some additional assumptions had to
be made, are in good agreement with ours.

\section{Discussion of the results}

\subsection{Attractor states}

The accessible region of the total pure state vector can be
decomposed into different zones, each of which yielding local states
with a common feature such as purity or energy probability distribution.
Provided the thermodynamic limit (which is defined below) applies to
the system, each accessible region is almost entirely filled with one
dominant zone, corresponding to one special local state. This (equilibrium) state can thus be
considered a local attractor. Hence even for trajectories starting in a
tiny ``non-equilibrium zone'', corresponding to some strongly
time dependent local state, after some time the gas system is most likely
to be found in the time independent attractor state, no matter where exactly the total
system keeps wandering around on its trajectory.\\
Those attractor states are as follows:\\
(i) Microcanonical conditions:
Here we have found
\be
\hat{\rho}^g_{mic.}\approx \sum_{A,a} \frac{W_A^g}{N_A^g}|A,a\>\<A,a| 
\ee
where $A$ labels the respective energy eigenspace, $a$ the different eigenstates
within an eigenspace, $W_A$ is the probability to find the system at
the energy $E_A$ and $N_A$ is the degree of degeneracy of the
corresponding energy level.\\
$\hat{\rho}^g_{mic.}$ is the state with the highest entropy that the
initial state can possibly evolve into under strict energy
conservation.\\
(ii) Canonical conditions: In this case we have shown that
\be
\label{canat}
\hat{\rho}^g_{can.}\approx \frac{e^{\frac{-\hat{H}^g}{kT}}}{\mbox{Tr}\lk e^{\frac{-\hat{H}^g}{kT}}\rk  }=\frac{\sum_{A,a}e^{-\frac{E_A}{kT}}|A,a\>\<A,a|}{\sum_A N_A e^{-\frac{E_a}{kT}}}
\ee
This state is
obviously the same one would have obtained, if one had applied Jaynes' principle,
taking energy as the only relevant observable.\\

\subsection{Quantum-thermodynamic limit}

For the above to hold, the system has to fulfill certain
requirements which then define the (quantum-) thermodynamic limit.\\
For both conditions the weak coupling limit should apply to the total
composite system, i. e.:
\be 
\hat{H}=:\hat{H}^g+\hat{H}^c+\hat{I}\quad \mbox{with} \quad \<\hat{I}\>\ll\<\hat{H}^g\>,\<\hat{H}^c\>.
\ee
where $\hat{H}^g$ and $\hat{H}^c$ are the local Hamiltonians of
the system and the environment, respectively, with
$\lb\hat{H}^g,\hat{H}^c\rb=0$, $\hat{I}$ is some sort of effective interaction. The expectation values are to be taken for
energy eigenstates of the local uncoupled systems. If the coupling
was too strong the notion of subsystems would become meaningless, if
the coupling was too weak, thermalization times may become extremely long.\\
For a system to fulfill microcanonical conditions there are
further requirements:
\be
\label{kommu1}
\lb\hat{H}^{g},\hat{I}\rb=0 \quad \lb\hat{H}^c,\hat{I}\rb=0 
\ee
If (\ref{kommu1}) holds, no energy is exchanged between the two
subsystems regardless of the strength of the interaction. Furthermore,
the following condition has to be met:
\be
\label{apr1}
\sum_B\frac{(W_B^c)^2}{N_B^c}\ll\sum_A\frac{(W_A^g)^2}{N_A^g}
\ee
This holds if the environment system, c,  occupies energy levels of higher degeneracy than the
considered system, g, and/or if its energy probability distribution is
broader. This is likely to be the case, if the environment is much
larger than the considered system.\\
For canonical conditions the full system state density $n(E)$ must be
large at
those energies that the full system occupies. In addition, the function $\frac{d}{dE^c}\ln n^c(E^c):=\frac{1}{kT}$ has to be
approximately constant over some energy range. Within this range
the state density of the environment $n^c(E^c)$ has to be higher than
the state density of the gas system $n^g(E^g)$ in the corresponding
range. All this is, again, typical for ``large systems'' and for the
container system being even larger than the gas system.\\
In this definition of the thermodynamical limit there is no
necessity for the systems to consist of many particles (cf.\cite{JEN85}); nevertheless,
the above mentioned criteria are most likely fulfilled by such
systems.\\

\subsection{Ergodicity}

In classical statistics ergodicity (of isolated systems) is meant to
imply hat the time average (of an individual system) equals the
ensemble average. Experimentally accessible is only the former, but
then it becomes questionable, under what conditions this equivalence
should actually hold. And why should the (statistical) entropy based
on the ensemble distribution have something to say about an individual system?\\
In the quantum treatment the reduced density operator $\hat{\rho}_g$
describes, indeed, the individual subsystem in interaction with its
individual environment. Obviously the
underlying quantum nature renders irrelevant questions like: How long
do we have to wait until the single system has actually visited all
those different pure states as required by the postulated
ensemble-properties.\\
Only if one attempts to find a ``quasi classical'' interpretation of
the reduced von Neumann entropy, the missing information about the
actual local state may be asribed to our ``ignorance''\\
The fact that an individual system (embedded in a quantum environment)
``simulates'' a whole ensemble \cite{ESP76} may be considered a
remarkable signature of ``quantum parallelism'' - a parallelism that
has been implicitly anticipated since the beginnings of statistical
mechanics.\\ 
This kind of ``ergodicity'' appears to be a natural ingredient of
quantum mechanical system-environment scenarios. While the local
entropy is a property of each individual total state vector, for this
vector itself ergodicity is not needed at all.\\

\section{concluding remarks}

In this paper we have considered Hilbert space in quantum mechanics as the
analog to phase space in classical mechanics. This analog is supported
by the fact that in both cases ``micro-states'' correspond to state
space points, and their dynamics to deterministic trajectories.\\
If the total system is partitioned into  two subsystems ( the
considered or ``gas'' system, g, and the environment or ``container'',
c) , the reduced
local state, $\hat{\rho}^g$, for the gas subsystem can be calculated
from the pure total system
state $|\psi\>$. Any such local state has typically non-zero entropy and
different total system states may very well yield the same
local system state. This fact, in turn, allows to define a probabilty
distribution of local entropies on the space of total pure
states. This distribution has been the main target of our
investigation.\\
For composite systems we have specified classes of interactions,
i. e., microcanonical (mic.) and canonical (can.) Hamiltonians,
respectively, which generate respective Hilbert space
trajectories. These trajectories cannot cross the boundaries of
different ``accessible regions'' which are set by the condition type
(``mic.'' or ``can.''). This, again, is reminescent of classical systems for which the
trajectories may be confined to
certain energy shells in phase space.\\
We have studied the fundamental mechanisms underlying
Jaynes' principle with respect to energy exchange. Jaynes' principle,
which may be taken as a rule for unbiased reasoning, thus coincides
with emergent physical laws. Similar
considerations should hold also for other interaction types like
particle exchange (grand canonical condition). In any case, quantum
mechanics should play a central role in the foundations of
thermodynamics, not only, e. g., in the low temperature limit. The predominance of
local equilibrium states does not imply that non-equilibrium states
were inaccessible or of minor interest. In fact any quantum machine,
in particular the quantum computer, will require significant
deviations from equilibrium. Machine design must specify means to
prevent the system from running into the typical equilibrium behaviour.

\section{Appendix}
\label{app}

The integrals that are to be evaluated in Sect. {\bf \ref{int}} are essentially of
the form:
\be 
Z=\int z_l^{u_l}(\{\phi_n\})z_m^{u_m}(\{\phi_n\})\det{\mathcal F}(\{\phi_n\})\prod_n^{d-1}d\phi_n
\ee
where the $u$ are integers and the $z$ are
parametrized by the $\phi$ to lie on a hypershere of radius $R$ and
dimension $d$. (The $z_l$ correspond to the real and imaginary parts of
the amplitudes, respectively, $d$ is the degree of degeneracy of a given subspace
and $R^2$ the probability of this subspace to be occupied, cf. Sect. {\bf \ref{lan}}). The $z_s$ are related to cartesian coordinates $x_s$ by:
\be 
\frac{z_s(\{\phi_n\})r}{R}=x_s
\ee
where $r$ is a radial variable. The technique to solve this integral,
$Z$, 
is basically the same as used to calculate the surface area of a
hypersphere of arbitrary dimension. This surface area is also the
special case of $Z$ for $u_l=u_m=0$.\\
Defining
\be
Z_1:=\int_o^{\infty}e^{-r^2}\left(\frac{r}{R}\right)^{u_lu_m+d-1}dr
\ee
the product $Z_2=ZZ_1$ may be written as:
\be
Z_2:=\int_o^{\infty}e^{-r^2}\left(\frac{z_lr}{R}\right)^{u_l}\left(\frac{z_mr}{R}\right)^{u_m}\left(\frac{r}{R}\right)^{d-1}\det{\mathcal F}\prod_n^{d-1}d\phi_ndr
\ee
Now, this is an integral over all space, written in angular and radial
coordinates that can be converted to cartesian coordinates yielding:
\be 
Z_2=\int e^{-\sum_s x_s^2} x_l^{u_l}x_m^{u_m}\prod_s^ddx_s
\ee
Since this integral factorizes completely, both $Z_2$ and $Z_1$ can be
evaluated using standard tables of integrals, and depend only on $R,d,u_l,u_m$. (Note that $Z_2$
vanishes if any of the $u$ is odd.) $Z$ is then found to be
\be
Z(R,d,u_l,u_m)=\frac{Z_2(R,d,u_l,u_m)}{Z_1(R,d,u_l,u_m)}
\ee
The averages that are to be computed in Sect. {\bf \ref{int}} are of the
form:
\be
A(R,d,u_l,u_m):=\frac{Z(R,d,u_l,u_m)}{Z(R,d,0,0)}
\ee
They are all invariant with respect to exchange of the $u$'s. Here we need only:
\be
\label{sym}
A(R,d,0,1)=A(R,d,1,1)=0,\quad A(R,d,0,2)=\frac{R^2}{d}
\ee
which could also have been found from simple symmetry considerations,
and
\be
\label{ku}
A(R,d,2,2)=\frac{R^4\Gamma\left(\frac{d}{2}\right)}{4\Gamma\left(\frac{d}{2}+2\right)}
\quad
A(R,d,0,4)=\frac{3R^4\Gamma\left(\frac{d}{2}\right)}{4\Gamma\left(\frac{d}{2}+2\right)}
\ee
All $A$'s are invariant with respect to exchange of the $u$'s. 

We thank A.~Otte, M.~Stollsteimer, F.~Tonner, M.~Michel, H.~Schmidt,
T.~Haury and P.~Borowski for
fruitful discussions.\\
Financial support by the Deutsche Forschungsgemeinschaft is gratefully acknowledged.


\begin{thebibliography}{9}
\addcontentsline{toc}{chapter}{Literaturverzeichnis}

\bibitem{NEU30} J.~v.~Neumann, {\em Zeitschr. f. Physik} {\bf 57}, 30 (1930)

\bibitem{BLA59} J.~M.~Blatt,  {\em Progr. Theor. Phys.} {\bf 22}, 745 (1959)

\bibitem{BER55} P.~G.~Bergmann, J.~L.~Lebowitz {\em Phys. Rev.} {\bf 99}, 578 (1955)

\bibitem{PEN70} O.~Penrose, {\em Foundations of Statistical Mechanics}
(Pergamon Press, New York, 1970)

\bibitem{TER91} Ya.~P.~Terlitskii, {\em Statistical Physics} (North Holland Publ., Amsterdam, 1991)

\bibitem{GEM01} J.~Gemmer, A.~Otte, G.~Mahler, {\em Phys. Rev. Lett.}
{\bf 86}, 1927 (2001)

\bibitem{BER94} J.~Berger, {\em Physics Essays} {\bf 7}, 281 (1994)

\bibitem{ZUR94} W.~Zurek, J.~ Paz,
{\em Phys.~Rev~Lett.} {\bf 72}, 2508 (1994), comment {\em Phys.~Rev.~Lett.} {\bf 75}, 351 (1995)

\bibitem{BAL99} L.~E.~Ballantine, {\em Quantum Mechanics} (World
Scientific, Singnapore, 1999).

\bibitem{LAN61} R.~Landauer,  {\em IBM J. Res. Develop.} {\bf 5}, 183 (1961)

\bibitem{MLI93} M.~Li, P.~Vitanyi, {\em An Introduction to Kolmogorov
Complexity and its Applications}, Chapter 8, (Springer, New York,
Berlin 1993).

\bibitem{JAY57} E.~T.~Jaynes, {\em Phys. Rev.} {\bf 106}, 620 (1957)

\bibitem{JAYWS} For a web-site dedicated to E.~T.~Jaynes, see {\em http://bayes.wustl.edu}

\bibitem{BLU81} K.~Blum, {\em Density matrix. Theory and applications}
(Plenum Press, New York, 1981)

\bibitem{GEM02} J.~Gemmer, G.~Mahler, {\em Eur. Phys. J. D} {\bf 17},
385 (2001) 

\bibitem{TAS98} H.~Tasaki, {\em Phys. Rev. Lett.} {\bf 80}, 1373 (1998)

\bibitem{TSA88} C.~Tsallis,  {\em J. Stat. Phys.} {\bf 52}, 479 (1988)

\bibitem{DIU89} B.~Diu, C.~Guthmann, D.~Lederer, B.~Roulet, {\em
Elements de Physique Statistique} (Hermann Editeurs des Sciences et des
Arts, Paris, 1989)

\bibitem{ALL00} A.~E.~Allahverdyan, Th.~M.~Nieuwenhuizen, {\em
Phys. Rev. Lett} {\bf 85}, 1799 (2000)

\bibitem{LUB78} E.~Lubkin,
{\em Math. Phys.} {\bf 19}, 1028 (1978). 

\bibitem{SCH89} E.~Schr\"odinger, {\em Statistical Thermodynamics} (Dover Publ.,New York, 1989).

\bibitem{JEN85} R.~V.~Jensen, R.~Shankar, {\em Phys. Rev. Lett.} {\bf 54}, 1879 (1985)

\bibitem{ESP76} In the older literatur this situation has been
characterized as an ``improper mixture'', cf., e.g., B.~d'Espagnat
{\em Conceptual Foundations of Quantum Mechanics} (Benjamin, London, 1976)


\end{thebibliography}
\end{document}